\documentclass[reprint,amsmath,amssymb,aps]{revtex4-2}

\usepackage{dcolumn}
\usepackage{bm}
\usepackage{amsmath}
\usepackage{amsfonts}
\usepackage{color}
\usepackage[dvipsnames]{xcolor}
\usepackage{graphicx,graphics}
\usepackage{epstopdf}
\usepackage{float} 
\usepackage{pifont}
\usepackage{etoolbox}
\usepackage{indentfirst}
\usepackage{braket}
\usepackage{ulem}
\usepackage{dsfont}
\usepackage{slashed}

\usepackage{bookmark,hyperref}
\hypersetup{
	colorlinks,
	citecolor=blue,
	filecolor=blue,
	linkcolor=blue,
	urlcolor=blue
}

\def\bb{\begin{equation}}
\def\ee{\end{equation}}

\def\5{\hspace*{5mm}}

\begin{document}

\title{Cosmological Bounces Induced by a Fermion Condensate}

\author{Giorgi Tukhashvili}
\email{giorgit@princeton.edu}
\author{Paul J. Steinhardt}
\email{steinh@princeton.edu}
\affiliation{Department of Physics, Princeton University, Princeton, NJ, 08544, USA}

\date{\today}

\begin{abstract}
We show that it is possible for fermion condensation of the Nambu-Jona-Lasinio type to induce a non-singular bounce  that smoothly connects a phase of slow contraction to a phase of expansion.  A chiral condensate -- a non-zero vacuum expectation value of the spinor bilinear  $\langle \bar{\Psi}\Psi \rangle$ -- can form spontaneously after a slow contraction phase smooths and flattens the universe and the Ricci-curvature exceeds a critical value.  In this approach, a high density of spin-aligned free fermions is not required, which avoids the problem of generating a large anisotropy and initiating chaotic mixmaster behavior during the bounce phase.

\end{abstract}

\maketitle


\noindent
{\it Introduction.}
The salient cosmological properties of the observable universe -- its homogeneity, isotropy, spatial flatness, and nearly scale-invariant spectrum of density perturbations -- can be explained by a period of slow contraction followed by a non-singular bounce that smoothly connects the contracting phase  to the current period of expansion \cite{Cook:2020oaj}.  During a slow contraction phase, the scale factor shrinks gradually as $a(t) \sim |t|^{1/\varepsilon}$ for $t \rightarrow 0^{-}$ where $\varepsilon \gg 3$, and the energy density driving the slow contraction grows quickly as $1/a^{2 \varepsilon}$. Slow contraction has been shown to be a  {\it robust} smoother, meaning that homogeneity, isotropy and flatness are achieved during slow contraction even if the initial state is far from smooth and flat \cite{Ijjas:2020dws};  it  is a {\it rapid} smoother, meaning that the smoothness is achieved after $a(t)$ shrinks by just  a few e-folds \cite{Ijjas:2021wml}, leaving ample time for the generation of a nearly scale-invariant spectrum of density fluctuations during the remaining slow contraction phase \cite{Ijjas:2020cyh,Ijjas:2021ewd}; and it is a {\it universal} smoother, meaning that the entire universe is smoothed and flattened even over distances that are  greater than a Hubble radius apart throughout the smoothing process \cite{Ijjas:2021gkf}. 

To ensure that these desirable features generated during the  contraction phase  survive into the expanding phase, a smooth transition is needed.  One approach is a non-singular {\it classical bounce} at an energy density sufficiently far below the Planck density that quantum gravity effects do not spoil smoothness. According to well-known singularity theorems \cite{Penrose:1964wq,Hawking:1966jv,Hawking:1970zqf}, a classical bounce requires a stable form of stress-energy that violates the null energy condition (NEC) or a modification of Einstein general relativity that avoids the null convergence condition (or both).  Although this comes at the cost of some form of  instability in certain cases \cite{Rubakov:2014jja}, examples that avoid all known instabilities have been identified ~\cite{Ijjas:2016tpn,Ijjas:2016vtq,Ijjas:2018cdm}. 

In this paper, we consider an alternative {\it semi-classical
bounce} induced by fermion condensation of the Nambu-Jona-Lasinio (NJL) type \cite{Peskin:1995ev,Weinberg:1996kr}.  The bounce is semi-classical in  that the  formation of the fermion condensate is a  quantum effect but its impact on cosmic evolution can be described by classical equations of motion.

To incorporate fermions in general relativity, Einstein's theory can be naturally extended to include torsion, where the spin of the matter fields is the source for torsion in much the same way as the matter fields are a source of curvature in Einstein's theory.  In the resulting Einstein-Cartan-Sciama-Kibble (ECSK) theory \cite{Sciama:1958we,Kibble:1961ba}, the torsion is non-dynamical, and so can be integrated out and traded in favor of spin-spin interactions.  If the torsion is sourced by the Dirac fermions,  the spin-spin interaction is  a {\it four-fermion  interaction} term $A_\mu A^\mu$ \cite{Weyl:1950xa,Kibble:1961ba}, where $A_\mu$ is the fermion axial current.

As shown here, the NJL chiral symmetry breaking phase transition can be triggered by  the gravitational background if the absolute value of the Ricci curvature rises above a critical value during slow contraction.  The background-triggered chiral symmetry breaking can occur naturally after the slow contraction phase smooths and flattens the universe but well before the energy density would reach the Planck density when quantum gravity effects become non-negligible.  At this point in time, the  $\langle A_{\mu}\rangle $ and the fermion vector current $ \langle V_{\mu}\rangle$ and, consequently, the spin and fermion density are all negligible, and they remain negligible throughout the bounce as well.  However, due to the breaking of chiral symmetry,  $\langle A_{\mu} A^{\mu}\rangle- \langle V_{\mu} V^{\mu}\rangle \sim \langle (\bar{\Psi}{\Psi})^2 \rangle$ becomes non-zero, producing a negative contribution to the Friedmann equation that grows rapidly enough to act as the NEC-violating interaction needed to induce a bounce.

The fact that the spin and fermion density remain negligible throughout the slow contraction and bounce phases is an important advantage compared to previous proposals for spin-induced bounces.
It was first noticed by Trautman \cite{Trautman:1973wy} that spin-spin interactions can act as an NEC-violating source that induces a bounce.  His treatment, like those of  later authors \cite{Alexander:2008vt,Magueijo:2012ug}, assumed a high density of spin-aligned fermions to obtain a non-zero average spin $\braket{{S^\rho}_{\mu \nu}}$ and, thereby, a non-zero value of the spin-spin interaction.
This approach is difficult to combine with slow contraction, though. If the large fermion energy density of physical particles is reached, it produces a non-zero spacelike value of $\langle A_{\mu}\rangle$ and spinor vector  current $\langle V_{\mu}\rangle$ that creates a large anisotropy, pushing  the universe away from Friedmann-Robertson-Walker (FRW) and triggering chaotic mixmaster behavior. This problem is avoided in the NJL approach considered in this paper.

\begin{figure}[b]
	\includegraphics[width=0.45\textwidth]{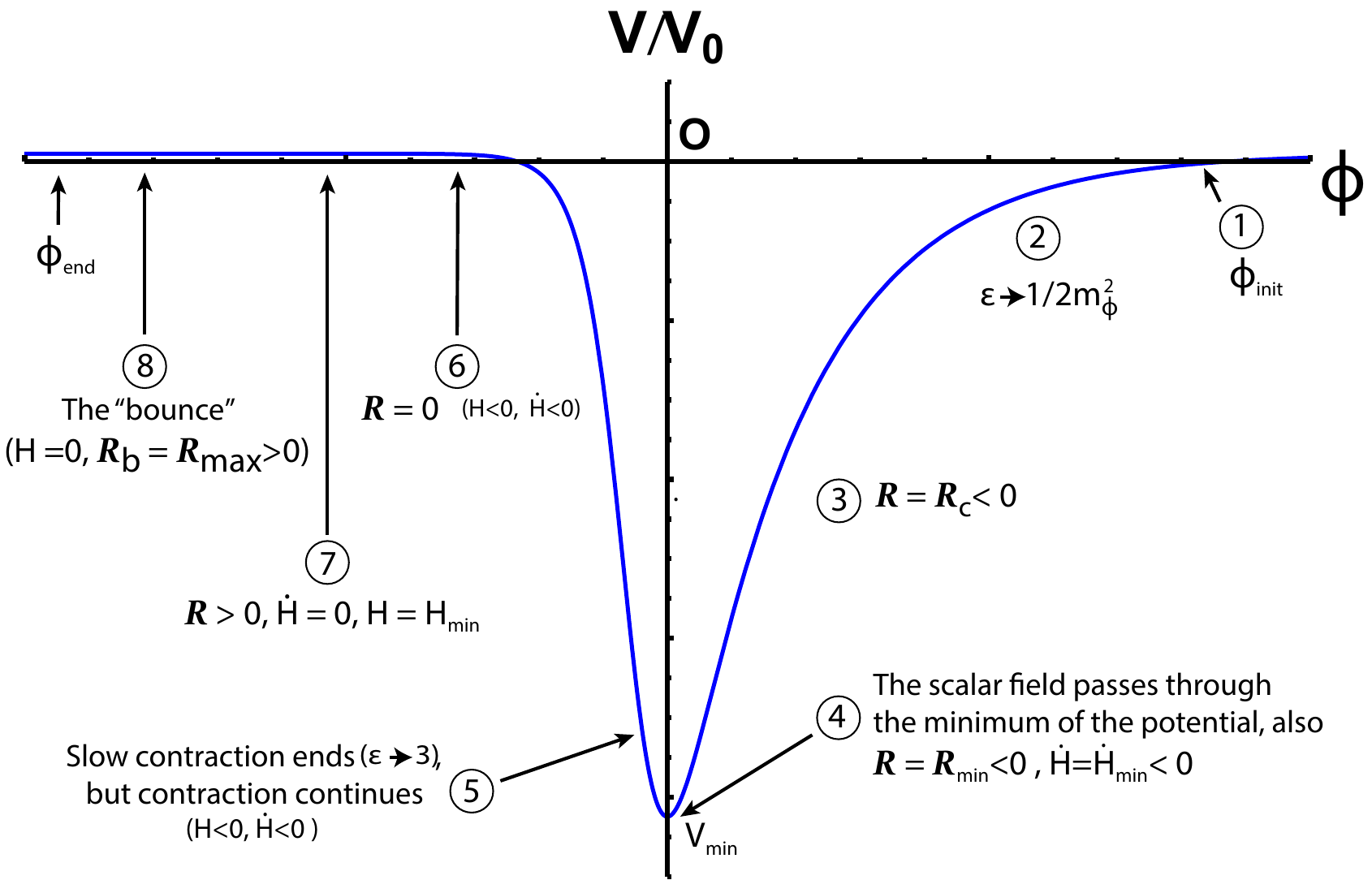}
	\caption{A qualitative sketch of the effective potential used in this paper and the various stages of evolution according to the equations of motion. The field begins at some $\phi=\phi_{\rm init}$ (stage 1) on the right hand side of the minimum $V(\phi_{\rm m}=0)$.  Evolving towards the left and down the exponentially steep potential triggers a phase of  slow contraction during which  the universe is rapidly smoothed and flattened and  $\varepsilon$ approaches $1/2 m_{\phi}^2 >3$ (stage 2).   At this point, $H$ is negative and decreasing.   As the field continues down the potential
the Ricci curvature $R$ (which is negative) falls below a critical negative value $R_c$ that switches on the NJL mechanism (stage 3).  As the field reaches the minimum of the potential, $R$ and $\dot{H}$ also reach their minimum values (stage 4). Slow contraction ends as $\phi$ crosses the minimum of the potential $V_{min}$,  but  ordinary contraction continues during stage 5 ($R$ and $\dot{H}$ are negative though increasing) as the field continues to roll up to the left side of the potential onto the plateau.  At some moment the energy of the condensate is large enough to flip the sign of the Ricci curvature (stage 6); and then flip the sign of $\dot{H}$ (stage 7); after which $H$ starts to increase.  Eventually, $H$ increases to the point where it crosses from negative to positive, corresponding to the bounce (stage 8).  Finally, due to the Hubble friction when $H>0$, the field comes to rest at some point $\phi_{end}$ along the positive plateau. }\label{fig1}
\end{figure}

\vspace{.1in}
\noindent
{\it Slow contraction via scalar fields.}
Beginning from a highly inhomogeneous, anisotropic, and spatially curved initial condition, a period of slow contraction can robustly, rapidly and universally smooth and flatten the universe \cite{Cook:2020oaj}.
A standard approach, which we incorporate here, is that  the slow contraction is achieved as a scalar field $\phi$ evolves down a steep negative exponential portion of its potential, $V(\phi) \approx - V_0 \, {\rm exp} (-\phi/m_{\phi})$ where $0<V_0 \ll1$ and  $m_{\phi} <1/\sqrt{6}$ in reduced Planck units ($8 \pi G=1$, where $G$ is the Newton gravitational constant).  An example is the right hand side of the potential shown in Fig.~\ref{fig1}. Beginning from inhomogeneous, anisotropic, and curved initial conditions that are far from FRW (stage 1 in Fig.~\ref{fig1}),  the universe rapidly smooths and flattens, approaching an FRW attractor solution with equation of state $\varepsilon = (3/2) (1 + p/\rho) = 1/(2 m_{\phi}^2 ) >  3$ (stage 2 in Fig.~\ref{fig1}).

The slow contraction phase ends when $\phi$ passes the minimum of the potential $V_{\rm min}$ and begins to roll up the potential towards a positive plateau (stage 5).  Contraction continues and the scalar field kinetic energy continues to increase due to Hubble anti-friction; however,  the contribution of $V(\phi)$ becomes negligible. causing the equation of state to approach $\varepsilon \rightarrow 3$ \cite{Kist:2022mew}.
For the example illustrated in this paper, we have chosen:
\begin{align}\label{scalar_pot}
	\nonumber V (\phi) = & V_{DE} - V_0 \left( 1 - \frac{m_*}{2 m_\phi} \right) \exp \left( - \frac{\phi}{m_\phi} \right) \times \\
	{} & \times \frac{\tanh \left( \frac{\phi }{m_*} \right) + 1}{ \left( 1 - \frac{m_*}{ m_\phi} \right) \tanh \left( \frac{\phi }{m_*} \right) + 1 }
\end{align}
which has an exponentially increasing form beyond the minimum with a different slope parameterized by $m_*$.
The field continues to evolve until it reaches the positive plateau on the other side of the potential minimum (stage 8).
(N.B. The pieced-together potential in Eq.~(1) is constructed for purposes of illustration and studied in detail in Ref.~\cite{Kist:2022mew} where it was shown to rapidly smooth and flatten spacetime beginning from initial conditions that are far from FRW.   
As shown in \cite{Kist:2022mew},  the same supersmoothing results can be achieved by using any negative potential satisfying $|V' /V| >5$, including ones of simpler form.  We use this example because it is well-documented, but our results are not sensitive to the choice in Eq.~(\ref{scalar_pot}).)

\vspace{.1in}
\noindent
{\it Fermions and the NJL-mediated bounce.}
The NEC violation and the non-singular bounce are entrusted to a Dirac spinor (spin-$1/2$ fermion).  The diffeomorphism group does not admit a spinorial representation, so we must use the first-order formalism of GR to couple spinors with gravity. In the first order formalism, the variables of the gravitational field are the tetrad, $e^I$, and the spin connection $\omega^{IJ}$. The metric tensor can be built from the tetrads: $g_{\mu \nu} = e^I_\mu e^J_\nu \eta_{IJ}$ \cite{footnote}. Finally, our model is based on the following action (the integration sign is implicit in front of each term):
\begin{align}\label{action_torsion_IO}
	\nonumber {} & \mathcal{S} = - \frac{1}{4} \epsilon_{A B M N}
	e^A \wedge e^B \wedge R^{M N} \\
    \nonumber {} & - \frac{i}{2 \cdot 3!} \epsilon_{A B M N} e^A \wedge e^B \wedge e^M \wedge \Big( \overline{\Psi} \gamma^N D \Psi - D \overline{\Psi} \gamma^N \Psi \Big)  \\
	{} & - \star 1
	\left[ \frac12 (\partial \phi)^2 + V (\phi)  
	+ m_{\Psi} \overline{\Psi} \Psi \right] \\
    \nonumber {} & +  \star 1 ~\xi
	\left[ \big( \overline{\Psi} \Psi \big)^2 + \big( i \overline{\Psi} \gamma_5 \Psi \big)^2 \right] .
\end{align}
The first line is the Einstein-Hilbert term; the second is the kinetic term for the massless Dirac spinor; the first two terms on the third line stand for the action of the scalar; while the very last term is the mass term for the spinor.  We   take $m_{\Psi}^4$ to be small compared to all other energy scales and densities during the bounce.

On the fourth line, we have the four-Fermi interactions generated by the torsion-spinor current couplings after the torsion is integrated out \cite{Paper2}. These four-Fermi interactions are invariant under the chiral transformations, $\Psi \rightarrow e^{i \alpha \gamma_5} \Psi$, and are equivalent to the Nambu-Jona-Lasinio model interactions. The volume form $\star 1 = - 1 / 4! \epsilon_{A B M N} e^A \wedge e^B \wedge e^M \wedge e^N = d^4 x \det (e)$. The connection $\omega^{IJ}$ is torsion free $D e^I = 0$.

In the NJL scenario considered here, there are no on-shell fermions present $ \braket{\overline{\Psi} \gamma^0 \Psi}=0$. Instead, the non-zero vacuum expectation value  of the scalar bilinear $ \braket{\overline{\Psi} \Psi}$ occurs through the NJL mechanism. The NJL gap equation with curvature and derivative corrections can be expressed as:
\begin{align}\label{gap_eq_orig}
	\nonumber {} & - \frac{1}{\Lambda^2} \frac{8 \pi^2}{\xi} u + \frac{\partial f_4 }{\partial u}  - \frac{2}{\Lambda^2} f_0 ~\nabla^2 u - \frac{1}{\Lambda^2} \left( \partial u \right)^2 \frac{\partial f_0 }{\partial u} \\
	{} & + \frac{R }{6 \Lambda^2} \left( 2 u f_1 + u^2 \frac{\partial f_1 }{\partial u} \right)
	  + \frac{2}{\Lambda^4} \mathcal{R}_1^2 \frac{\partial f_1 }{\partial u}  \\
	\nonumber {} & - \frac{2}{\Lambda^4} \mathcal{R}_2^2 \frac{\partial f_2 }{\partial u}  + \frac{2}{\Lambda^4} \mathcal{R}_3^2 \frac{\partial f_3 }{\partial u} = 0 ,
\end{align}
where $u \equiv - 2 \xi \braket{\overline{\Psi} \Psi} / \Lambda$, $\Lambda$ is the UV cut-off and the $\mathcal{R}_i^2$ are:
\begin{align} \label{gap_eq_orig1}
	 \nonumber \mathcal{R}_1^2 \equiv & - \frac{1}{120} \nabla^2 R + \frac{1}{288} R^2
	- \frac{1}{180} R_{\mu \nu} R^{\mu \nu} \\
	 {} & - \frac{7}{1440} R_{\mu \nu \rho \sigma} R^{\mu \nu \rho \sigma} ; \\
	\nonumber \mathcal{R}_2^2 \equiv & \frac{1}{12} \nabla^2 R + \frac{1}{36} R_{\mu \nu} R^{\mu \nu}
	+ \frac{1}{144} R_{\mu \nu \rho \sigma} R^{\mu \nu \rho \sigma} ;\\
	\nonumber \mathcal{R}_3^2 \equiv & \frac{1}{10} \nabla^2 R + \frac{1}{72} R^2
	+ \frac{7}{180} R_{\mu \nu} R^{\mu \nu} + \frac{1}{60} R_{\mu \nu \rho \sigma} R^{\mu \nu \rho \sigma} ;
\end{align}
where
\begin{align} \label{gap_eq_orig2}
	\nonumber f_1 \equiv & \frac{1}{1 + u^2} - \frac{y^2}{y^2 + u^2}
	- \log \left( \frac{1 + u^2}{y^2 + u^2} \right) ; \\
	\nonumber f_2 \equiv & \frac{1}{2 ( 1 + u^2 )^2} - \frac{y^4}{2 ( y^2 + u^2 )^2} ; \\
	f_3 \equiv & \frac{1}{3 ( 1 + u^2 )^3} - \frac{y^6}{3 ( y^2 + u^2 )^3} ; \\
	\nonumber f_4 \equiv & (1 - y^2) u^2
	+ \log \left( 1 + u^2 \right) \\
	\nonumber {} & - y^4 \log \left( 1 + \frac{u^2}{y^2} \right)
	- u^4 \log \left( \frac{1 + u^2}{y^2 + u^2} \right) ; \\
	\nonumber f_0 \equiv & f_1 + 4 f_3 +
	8 u^2 \left( \frac{1}{(1+u^2)^3} - \frac{y^4}{(y^2+u^2)^3}  \right) .
\end{align}
Eq. (\ref{gap_eq_orig}) is derived from the one-loop effective action, which itself is derived from the spinor part of Eq.~(\ref{action_torsion_IO}) in a curved background space-time
(see appendix of \cite{Paper2} for detailed discussion). Although $u$ is a scalar, its effective action induces derivative interactions essential for violating the null energy condition, which does not occur for canonical scalar fields.
We used the Riemann Normal Coordinate method to derive the curvature corrections, while the derivative corrections were computed using the algorithm described in \cite{Gaillard:1985uh}. Eq. (\ref{gap_eq_orig}) is easily verifiable in the $\Lambda \rightarrow \infty$ limit; see, for example, \cite{Parker:2009uva} chapter 5.7.  Briefly,
the first two terms on line one represent the usual NJL gap equation with an invariant UV cut-off $\Lambda$ \cite{Nambu:1961tp}. The last two terms on line one are derivative corrections that arise when $u$ is time-dependent, non-uniform, or both. Lines two and three represent the curvature corrections discussed in Ref.~\cite{Paper2}. The ratio $y \equiv  2 m_{\Psi} / \Lambda \ll 1$, a dimensionless measure of  the explicit chiral symmetry breaking scale, sets the IR cut-off  necessary to avoid loop divergences that arise in the limit of exactly massless spinors.
The generalization of (\ref{gap_eq_orig}), with $\Psi$ having $N$ colors or flavors is straightforward. The large $N$ limit also removes the ambiguity between choosing $\braket{\overline{\Psi} \Psi}^2$ and $\braket{\overline{\Psi} \Psi \overline{\Psi} \Psi}$, as the difference between these two scales as $N^{-2}$.

In the absence of gravity, the fate of the condensate is determined by the NJL coupling strength $\xi$. If $\xi > \xi_c$, with $\xi_c = 2 \pi^2 / \Lambda^2$, then the chiral symmetry is spontaneously broken and $ \braket{\overline{\Psi} \Psi} \neq 0$. In the presence of gravity, the fate of  the condensate depends on both the NJL coupling and the Ricci curvature. In particular, solving  Eq.~(\ref{gap_eq_orig}) shows that large negative Ricci curvature favors chiral symmetry breaking and condensate formation.

\begin{figure}[t]
		\includegraphics[width=0.45\textwidth]{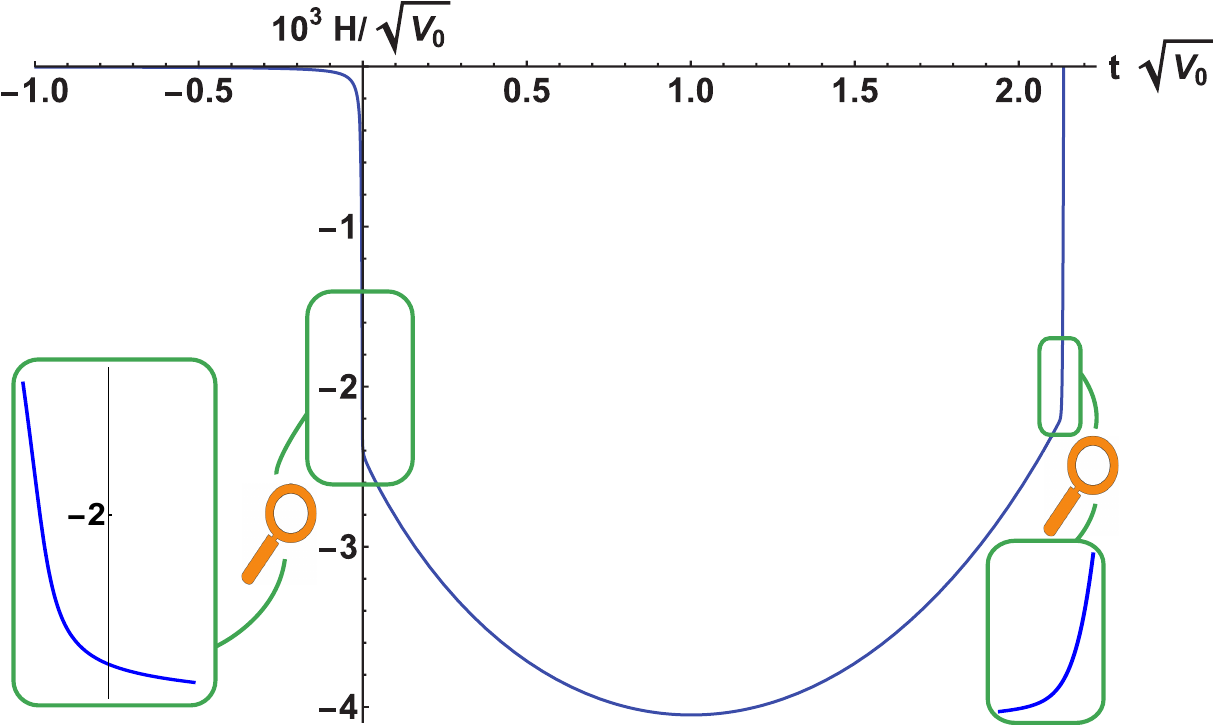}
	\caption{The normalized Hubble parameter $H/V_0^{1/2}$ as a function of the normalized time $
	V_0^{1/2} t$, where $t=0 $ corresponds to the moment when a  singularity (crunch) would have occurred if the fermions were absent and the potential were purely a negative exponential, $V(\phi) = - V_0 \exp \left( - \phi / m_\phi \right)$.
	  The cosmological  solution of interest was derived according to Eqs.~(\ref{gap_eq_orig}), (\ref{scalar_in_FRW}), (\ref{bounceeq}) and the potential in Eq. (\ref{scalar_pot}) and Fig.~ \ref{fig1}. The insets on the left and right show that apparently sharp features in the evolution of $H$ are actually smooth transitions. }
	\label{hubble}
\end{figure}

\begin{figure}[t]
		\includegraphics[width=0.45\textwidth]{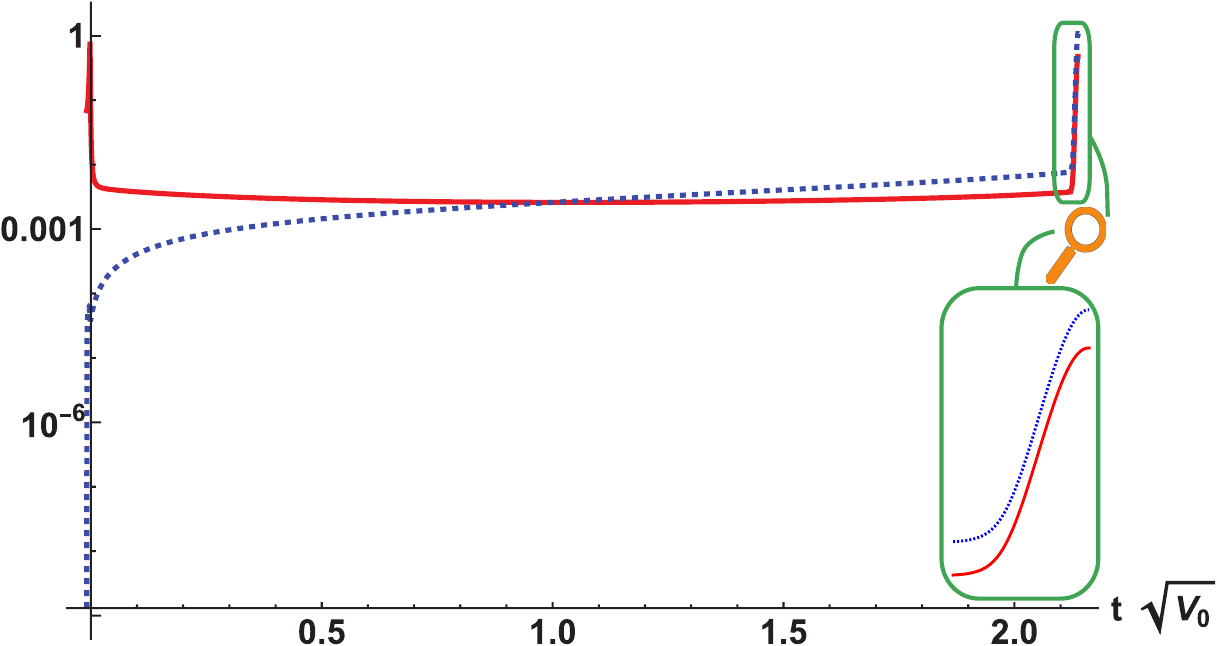}
	\caption{The behavior of the negative and positive contributions to $\dot{H}/V_0$ in eq. (\ref{hdoteq}):  solid red, the absolute value of the negative contributions (first two terms on the RHS of eq. (\ref{hdoteq})); dashed blue, the positive contribution (last term on the RHS). }\label{h_dot_fig}
\end{figure}

For example, if the constant  $\xi$ is below its critical value  and the universe begins to slowly contract, the Ricci curvature $R$ is negative and growing increasingly negative.  Above some critical value $0>R> R_c$,  the Ricci curvature is not sufficiently negative to break the chiral symmetry, so
$\braket{\overline{\Psi} \Psi} = 0$ and only the scalar field contributes to the Einstein equations.  Once the universe is sufficiently smoothed and flattened by slow contraction to be well-approximated as FRW (by stage 2 in Fig.~\ref{fig1}), the scale factor evolves as a power law in time, while the scalar field evolves as a logarithm:
\bb\label{log_scalar}
 a(t) = a_0 |t|^{2 m_{\phi}^2}; ~~
~ \phi (t) = m_{\phi} \log\left[ \frac{V_0 ~ t^2}{2 m_{\phi}^2 - 12 m_{\phi}^4 } \right],
\ee
at $t \rightarrow 0^{-}$.
The Ricci curvature is negative and decreasing as $R \propto - 1/ t^2$.  The Hubble parameter $H$ is also negative and decreasing as $2 m_{\phi}^2/t$.

Provided $\xi$ is not too far below $\xi_c$, there can come a time $t_c$ when the Ricci curvature $R$ falls below the critical value $R_c$ (stage 3 in Fig.~\ref{fig1}), while $R/\Lambda^2$ remains small enough for the curvature expansion in (\ref{gap_eq_orig}) to be valid. At that moment, a gravitational switch is thrown:  the chiral symmetry is broken and  the effective potential of the spinor bilinear develops energetically favorable minima away from  $u = 0$. As a consequence the scalar bilinear  acquires a non-zero vacuum expectation value,  $\braket{\overline{\Psi} \Psi} \neq 0$, and the equations in this FRW limit become:
\bb\label{scalar_in_FRW}
\ddot{\phi} + 3 \frac{\dot{a}}{a} \dot{\phi} + \frac{d V}{d \phi}  = 0 ;
\ee
\bb\label{bounceeq}
3 H^2 = \frac12 \dot{\phi}^2 + V + m_{\Psi} \braket{\overline{\Psi} \Psi}
- \xi \braket{\overline{\Psi} \Psi}^2 ;
\ee
\bb\label{hdoteq}
\dot{H} = - \frac12 \dot{\phi}^2 + \frac{m_\Psi}{6 H } \frac{d \braket{\overline{\Psi} \Psi}}{d t}
 - \frac{\xi}{3 H} \braket{\overline{\Psi} \Psi} \frac{d \braket{\overline{\Psi} \Psi}}{d t}  ;
\ee
while $\braket{\overline{\Psi} \Psi}$ evolves according to Eq. (\ref{gap_eq_orig}). The linear and quadratic curvature corrections in the gap equation (\ref{gap_eq_orig}), arise from the loop corrections and compared to the Einstein-Hilbert term, are $\hbar$ suppressed. This is the reason they are not included in eq. (\ref{bounceeq}).

The sign in front of the bilinear squared in the Friedmann equation, Eq.~(\ref{bounceeq}), corresponds to a   source of NEC violation.  Following the field equations, the Ricci curvature $R$ begins to grow, crossing from negative to positive (stage 6 in Fig.~\ref{fig1}), and $\dot{H}$ also increases until it flips from negative to positive (stage 7).  Once that occurs, which is well before $|H|$ reaches the Planck scale,  $H$ (which is still $<0$ and had been decreasing) begins to increase.  Finally, as field reaches the positive plateau, the bounce occurs: $H$ crosses from negative to positive and $a(t)$ reaches its minimal value.
At the same time, the Ricci curvature reaches its maximum value during this entire period of contraction, $R_b \approx 3.5 \times 10^{-4} = 3.5 \times 10^{-3} \Lambda^2$, which is sufficiently below the Planck scale to justify neglecting quantum gravity effects and higher order curvature corrections not included in Eq.~(\ref{gap_eq_orig}). 

\begin{figure}[b]
		\includegraphics[width=0.45\textwidth]{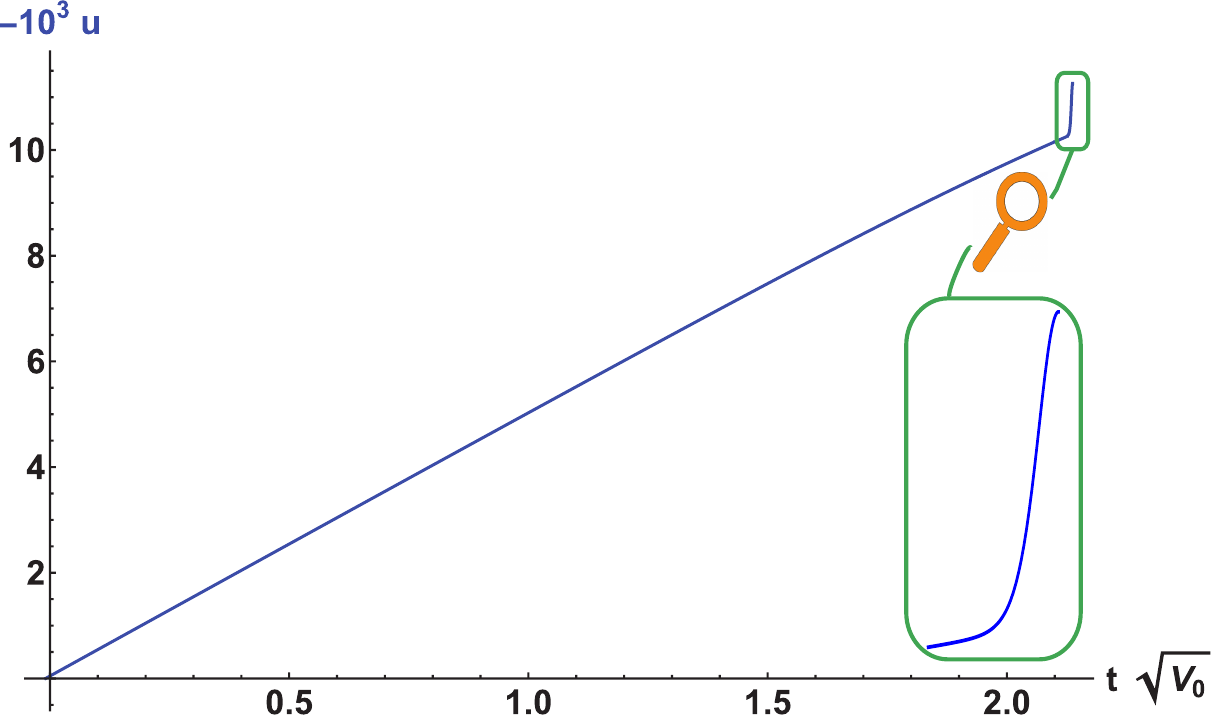}
	\caption{The chiral condensate $u = - 2 \xi \braket{\overline{\Psi} \Psi} / \Lambda$ as a function of the normalized time $V_0^{1/2} t$. At the moment of bounce (far right in the figure), the condensate reaches its maximum value, $\dot{u} (t_b) = 0$, as shown in the inset. This coincides with $R$ reaching its maximum value and $H$ passing through zero from negative to positive values.}\label{cond}
\end{figure}

Figs. \ref{hubble}, \ref{h_dot_fig} and \ref{cond} illustrate numerical solutions of the Hubble parameter, positive and negative contributions to $\dot{H}$, and the condensate, respectively, for an example where
$\Lambda = 1/ \sqrt{10}$; $m_{\phi} = 1/800$, $m_* = 1/2500$, $m_{\Psi} = \Lambda /200$ , $V_0 = 10^{-4}$, $V_{DE} = 0$, and $\xi = 0.99999 \xi_c$.  We set the  conditions when the chiral symmetry is first broken (Stage 3: $\sqrt{V}_0 t_c \approx - 0.009$) to be : $a(\sqrt{V}_0 t_c) = 1$, $u(\sqrt{V}_0 t_c) = 0$, $u'(\sqrt{V}_0 t_c) = - 5 \times 10^{-3}$.
The latter is set by the $m_{\Psi}$, as this scale is responsible for the explicit chiral symmetry breaking.

Fig. \ref{hubble} shows that $H(t)<0$ decreases rapidly as the scalar field rolls downhill ($t \lesssim 0$); decreases more slowly after the scalar field passes the minimum and rolls uphill; then increases slowly as $\dot{H}$ first increases above zero; and finally increases sharply and passes through zero on the right as the condensate begins to sharply rise near the bounce, (as illustrated in the inset in Fig.~ \ref{cond}).

Fig. \ref{h_dot_fig} shows the evolution of the positive and negative contributions to $\dot{H}$.  Soon after the fermion condensate forms, the positive contribution increases sufficiently fast to overtake the negative contributions, causing $\dot{H}$ to become positive, indicative of NEC violation.

The behavior of the condensate $u$ in Fig.~ \ref{cond} shows the evolution between the initial chiral symmetry breaking at $t=t_c$  (Stage 3 in Fig.~\ref{fig1})  and the bounce ($t=t_b$).   Before chiral symmetry breaking, the potential for $u$ has a single (global) minimum at $u=0$. When the chiral symmetry is broken, $u=0$ becomes a local maximum and there are global minima at $u=u_{\rm min}(t)$ on either side of the $u=0$.  The values of $u=u_{\rm min}(t)$ and the shape of the condensate potential vary with the Riemann curvature and, depending on parameters, may go through various stages of symmetry restoration and breaking.  The average effect is to drive $u$ farther from the $u=0$.
The key moment occurs just as the bounce approaches.  By this point, the value of $u$ is moving farther and farther from zero and lies beyond the minimum of the potential. As the bounce approaches, the curvature increases, the symmetry is restored,  and the potential away from $u=0$ becomes steeper and steeper.  The field, which had been moving away from $u=0$, is thereby forced to slow down and eventually reverse course.  The bounce corresponds to the moment when $u$ reaches its greatest magnitude ($\dot{u}_{b}=0$) and $R$ reaches its maximum value.

\vspace{.1in}
\noindent
{\it Summary and Outlook.}
In this paper, we have introduced the possibility of a semi-classical, non-singular, fermion condensate induced cosmological bounce mediated by the NJL mechanism.  The NJL mechanism does not require a high density of fermions or a violation of isotropy.
Instead, the non-zero vacuum expectation of the scalar bilinear occurs because the external gravitational field (more specifically, the Ricci curvature) acts as a switch for chiral condensate formation. The scenario naturally combines with a preceding phase of slow contraction that smooths and flattens the universe. 

We have stopped the evolution at the bounce ($H=0$). The stability of the NJL-mediated bounce to matter and gravitational perturbations; the reheating after the bounce, and the generation of nearly scale-invariant density perturbations will be the subject of forthcoming work.


\vspace{.1in}
\noindent
{\it Acknowledgements.}
	We thank A. Ijjas, G. Gabadadze, R. Woodard, D. Shlivko, and S. Alexander for many useful discussions and T. Damour for correspondence.  This work is supported in part by the DOE grant number DEFG02-91ER40671 and by the Simons Foundation grant number 654561.



\end{document}